\documentclass{sig-alternate-05-2015}

\usepackage[utf8]{inputenc}
\usepackage{tabularx}
\usepackage{booktabs}
\usepackage{subcaption}

\newcommand{\citep}[1]{\cite{#1}}
\newcommand{\citealp}[1]{\cite{#1}}
\newcommand{\citeyearpar}[1]{\cite{#1}}
\newcommand{\citet}[1]{\cite{#1}}

\begin{document}






%

\title{On the Ubiquity of Web Tracking:\\Insights from a Billion-Page Web Crawl}
%
%
%
%
%

\numberofauthors{2} 
%
\author{
%
%
\alignauthor
Sebastian Schelter\\
  \affaddr{Technische Universität Berlin}\\
  \email{sebastian.schelter@tu-berlin.de}
\alignauthor
Jérôme Kunegis\\
\affaddr{University of Koblenz--Landau}\\
  \email{kunegis@uni-koblenz.de}
}

\maketitle

\begin{abstract}
We perform a large-scale analysis of third-party trackers on the World Wide Web. We extract third-party embeddings from more than 3.5~billion web pages of the CommonCrawl 2012 corpus, and aggregate those to a dataset containing more than 140 million third-party embeddings in over 41 million domains. To the best of our knowledge, this constitutes the largest empirical web tracking dataset collected so far, and exceeds related studies by more than an order of magnitude in the number of  domains and web pages analyzed.

Due to the enormous size of the dataset, we are able to perform a large-scale study of online tracking, on three levels: (1)~On a global level, we give a precise figure for the extent of tracking, give insights into the structural properties of the `online tracking sphere' and analyse which trackers (and subsequently, which companies) are used by how many websites. 
(2)~On a country-specific level, we analyse which trackers are used by websites in different countries, and identify the countries in which websites choose significantly different trackers than in the rest of the world. 
(3)~We answer the question whether the content of websites influences the choice of trackers they use, leveraging more than ninety thousand categorized domains. In particular, we analyse whether highly privacy-critical websites about health and addiction make different choices of trackers than other websites. 

Based on the performed analyses, we confirm that trackers are widespread (as expected), and that a small number of trackers dominates the web (Google, Facebook and Twitter).  In particular, the three tracking domains with the highest PageRank are all owned by Google.  The only exception to this pattern are a few countries such as China and Russia. Our results suggest that this dominance is strongly associated with country-specific political factors such as freedom of the press. Furthermore, our data confirms that Google still operates services on Chinese websites, despite its proclaimed retreat from the Chinese market. 
We also confirm that websites with highly privacy-critical content are less likely to contain trackers (60\% vs 90\% for other websites), even though the majority of them still do contain trackers. 
\end{abstract}

\section{Introduction}
The ability of a website to track the pages read by its visitors has been present since the beginnings of the World Wide Web. In recent years however, another tracking mechanism has become widespread: that of third-party websites embedded into the visited site by mechanisms such as JavaScript and images. 
In fact, the majority of websites contain third-party content, i.e., content from another domain that a visitor's browser loads and renders upon displaying the website. 
Such an embedding of third-party content has always been possible, but was relatively rare, since most embedded images were located on the same server as the page itself, and in any case, the embedding of content was not intended for tracking.  

With the rise of social media and \emph{Web~2.0}, websites increasingly began to embed links (using various technologies) to third-party content, and thereby allowed the providers of such content to track users on a wide scale. Even if tracking is not the goal of a particular technology, the inclusion of third-party content occurs for a variety of reasons, e.g., advertising, conversion tracking, acceleration of content loading, and provision of widgets. Social media and advertising companies primarily use the data collected via these tracking mechanisms to improve their capability to show personalized, tailored advertisements to their users, which represent their main source of income. 

Regardless of their primary purpose, third-party components can (and in many cases do) track web users across many sites and record their browsing behavior. Therefore, these third-parties constitute a privacy hazard in many aspects, such as the collection of data about health conditions \citep{EFFHealth,Libert2015}, news consumption \citep{Trackography}, or their instrumentalization for mass surveillance by intelligence agencies \citep{TheIntercept}. In order to understand and control these hazards, it is desirable to gain a deeper understanding of the `online tracking sphere' as a whole. Previous research has studied small samples of this sphere, e.g., 1,200 English-language domains from Alexa's popular sites \citep{Krishnamurthy2009}, while researchers have only recently started to study larger datasets, e.g., tracking on the Alexa top 1 million domains \citep{Libert2015exposing}. The main reason for this is that, until lately, such a study would only have been possible for large companies possessing their own `copy' of the web. Recent developments however allow us to study this online tracking sphere at a large scale: the availability of enormous web crawls comprised of hundreds of terabytes of web data \citep{Spiegler2013}, such as CommonCrawl 2012.\footnote{\url{http://blog.commoncrawl.org/2012/07/2012-crawl-data-now-available/}} We process all 3.5 billion webpages from this corpus (which amounts to more than 200 terabytes of data), and extract the bipartite `tracking graph', which describes the tracking of more than 41 million pay-level domains by 355 tracking services. A pay-level domain (PLD) is a sub-domain of a public top-level domain that users pay for individually\footnote{\url{http://webdatacommons.org/structureddata/vocabulary-usage-analysis/}}. Subsequently, we use this data to study three key research questions: (i)~Which are the predominant tracking companies on the Web, and how many websites do they track? (ii)~How does the distribution of web trackers vary from country to country? How is that variation related to different political, socio-cultural and economic factors? (iii)~How are trackers distributed with respect to highly privacy-critical topics such as health and addiction? Do authors of such websites avoid web trackers?
  
To the best of our knowledge, the data extracted in the paper constitutes the largest empirical dataset of web trackers collected so far, containing an order of magnitude more domains than comparable studies. In this article, we extend our findings from previous work~\citep{Schelter2016} and provide the following contributions. (1)~We conduct the first analysis of tracking in the long tail of domains on the web, looking into 41~million pay-level domains, thereby surpassing previous work by three orders of magnitude, cf.~Section~\ref{chap:global}. We identify global structural properties of the online tracking sphere, e.g.,~we confirm the extraordinary tracking capability of Google Analytics (cf.\ \citealp{Roesner2012,Krishnamurthy2006}).  (2)~We uncover country-specific as well as category-specific patterns in the relations between third-party trackers by clustering their co-occurrences. (3)~We analyze the tracking company distribution in top-level domains (TLDs) of non-English speaking countries, and find that a small set of US-based companies have a dominating role in the majority of countries; a correlation analysis suggests that this role is strongly associated with political factors such as freedom of the press. Furthermore, our data confirms the finding that Google still operates tracking services on Chinese websites, despite its 
proclaimed retreat from the Chinese market~\citep{TheGuardianGoogleChina}. 
(4)~We compare the tracking capabilities on more than ninety thousand highly privacy-critical and less privacy-critical domains, finding that overall, websites with highly privacy-critical content do avoid trackers as opposed to other websites, even though the majority of them still do contain trackers.  We also show that certain types of trackers are in fact present with higher probability on websites with privacy-critical content, indicating a lack of awareness of the tracking ability of their underlying services. 
(5)~Finally, we provide our dataset to the scientific community\footnote{\url{https://ssc.io/trackingthetrackers}} and contribute it to the \emph{Koblenz Network Collection}~\citep{Kunegis2013}\footnote{\url{http://konect.uni-koblenz.de/networks/trackers-trackers}}.
\\
\\
\\
\\
\\

\section{Online Tracking Fundamentals}\label{sec:fundamentals}

In this section, we give a technical introduction to web tracking, and highlight resulting privacy implications.

\subsection{Technical Foundations}\label{sec:foundations} 

In its basic form, online tracking involves two actors: a \emph{user} browsing the Web, and the \emph{website} that she intentionally visits.  In addition to these, one or more services may be present that record her browsing to the website; we call such services \emph{third-parties}. 

We refer to third-parties as \emph{trackers} if their main purpose or the business model of their owning company depends on collecting browsing data of users. Therefore, we categorize advertising services and social network plugins as trackers, because their main source of income comes from targeted, personalized advertisements which heavily depend on user data. On the other hand, we do not consider content delivery networks as trackers, because their main business is to accelerate website loading (see~\citealp{Krishnamurthy2009,Englehardt16}).

Specifically, the user visits a website, which means that her browser issues an HTTP ``GET'' request to the web server hosting the website. The web server returns an HTTP response containing the HTML code of the website to render. This returned HTML code typically contains references to external resources, such as style sheets, JavaScript code and images that are required to render the page in the client browser. Next, the user's browser automatically issues requests for these additional resources.  

\begin{figure*}
  \centering
  \includegraphics[scale=0.9]{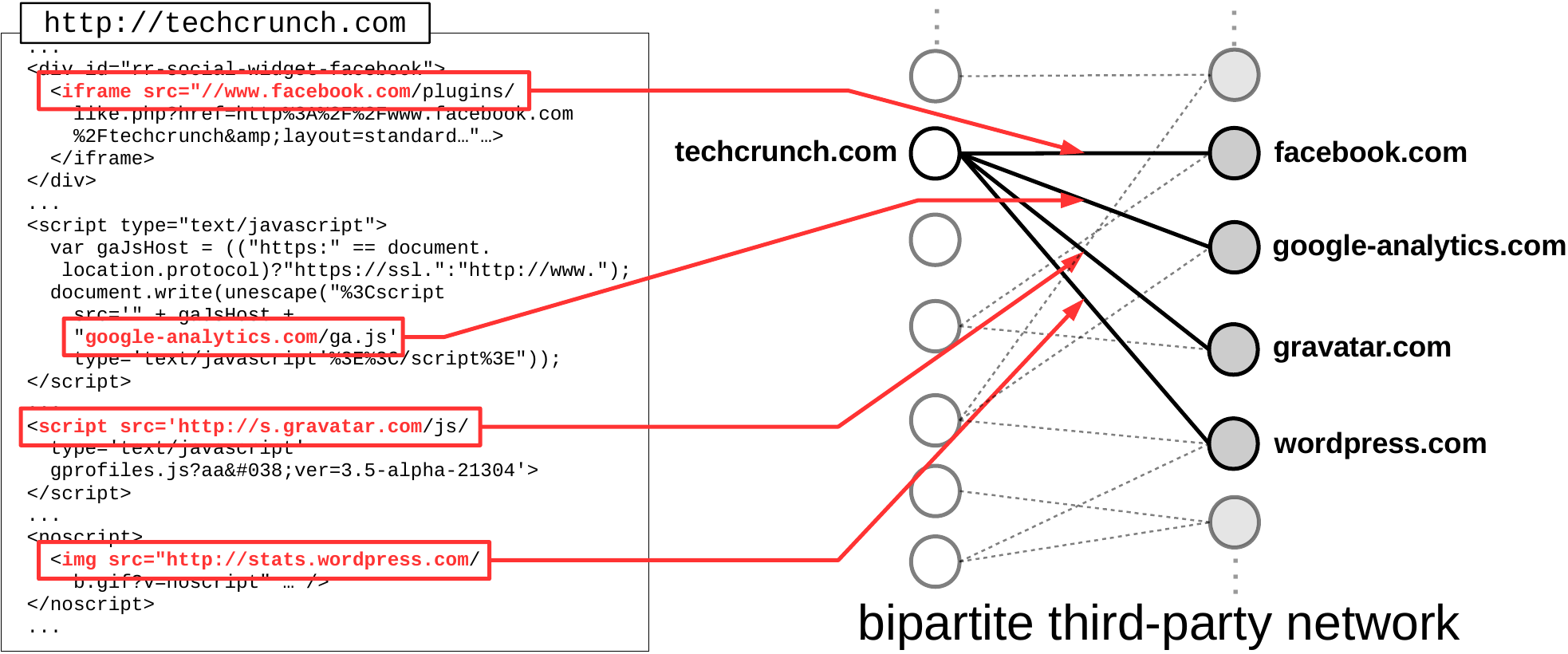}
  \caption{
    Example of our extraction process which results in the bipartite third-party network. We parse the HTML code of a website to extract the pay-level domains of all embedded third-parties. The captured embeddings form edges in the bipartite third-party network.
  }
  \label{fig:extraction}
\end{figure*}

In many cases, resources used by a web page reside on a different server than the one hosting the website. This possibility allows third-parties to track the user by recording and inspecting the external requests to their servers.

In the case of online tracking, the website's HTML code embeds external resources from a third-party which aims to track the user. A typical example for such an external resource is a piece of JavaScript code, which the user's browser will automatically load and execute from the third-party server. This external loading enables the third-party to record a wide variety of information about the user: (i) The third-party sees the current IP address of the user machine, which reveals the internet service provider as well as the approximate geolocation of the user. The browser typically announces its version, the underlying operating system, the user's screen resolution and other information which allow the third-party to compute a `fingerprint' of the browser. These fingerprints have been found to be sufficient to recognize individual users with a high precision \citep{Eckersley2010}. (ii)~The third-party has access to the URI of the page which the user is currently visiting, and, by looking at the HTTP referrer, also learns the URI of the previous page the user has visited. (iii) The third-party is able to read existing cookies designated for its domain and can set new cookies. If the user chose to be persistently logged in to a service such as Google, Facebook or Twitter, these third-parties will recognize the user as they will receive the persistent login cookies. 

Third-parties are either embedded dynamically such as via JavaScript or the \texttt{iframe} element, or statically via \texttt{link} or \texttt{image} tags. In the latter case, they lose some of their tracking abilities such as reading the referrer and browser properties such as the screen resolution.

\newpage
\subsection{Privacy Implications} 
Third-parties vary considerably.
They include advertisers, analytics services, social widgets, content delivery networks and image hosters. All of these have legitimate uses such as monetization, webpage optimization, A/B testing, conversion tracking, the provision of widgets, and fraud detection. However, the ability of many third-parties to record large portions of the browsing behavior of many users across a huge number of sites on the Web poses a privacy risk, and is the subject of ongoing legal 
disputes \citep{TheGuardian}. The data recorded by this tracking infrastructure has been reported to contain large portions of online news consumption \citep{Trackography}, as well as intimate, health-related personal data \citep{EFFHealth,Libert2015}, and personal data from social networks \citep{Mayer2012,Krishnamurthy2009OSN}. A large variety of browser-addons, (e.g., \citealp{Disconnect,Ghostery,PrivacyBadger}) have been developed to allow individuals to block online trackers from their computer. 

The ability to consume news and form a political opinion in an independent and unwatched manner, as well as the privacy of personal health-related data are vital for an open society, and should not be subject to commercially motivated data collection. Furthermore, recent reports suggest that intelligence agencies piggyback on online tracking identifiers to build databases of the surfing behavior of millions of people \citep{TheIntercept}. Frighteningly, this data collection seems to take place outside of the legal supervision of the governments of many of the people affected.

\section{Data Acquisition}
\subsection{Collection Methodology and Limitations} 

CommonCrawl~2012 is a very large publicly available crawl of the web, which consists of more than 3.5~billion HTML pages and amounts to approximately 210~terabytes in uncompressed form \citep{Spiegler2013}. While the CommonCrawl~2012 dataset is four years old as of 2016, we choose it over latter corpora, as it produces a more faithful representation of the underlying link structure of the web. This is due to the fact that it has been created using a breadth-first search crawling strategy \citep{Lehmberg2014}, which leads to a more realistic representation of the connected network structure of the web compared to the crawling strategy employed in latter corpora. In these latter corpora, the crawling process fetched domains from predefined lists and did not follow links, which results in a network with an unreasonably sized largest strongly connected component, which is not well suited for structural analysis.\footnote{\url{http://webdatacommons.org/hyperlinkgraph/2014-04/topology.html}}

During our extraction process, we represent websites and third-parties by their pay-level domains, basing our data acquisition strategy on the fact that the URLs of third-party services have to be embedded in the HTML of the websites. Our extractor takes an HTML document as input and retrieves all embedded domains of third-party services, as shown in Figure~\ref{fig:extraction}. In a first pass through the document, we parse the HTML and investigate the \texttt{src} attribute of \texttt{script}, \texttt{iframe}, \texttt{link} and \texttt{image} tags. In order to also find third-parties that are embedded via JavaScript code, we run a parser\footnote{\url{https://github.com/google/closure-compiler}} on all JavaScript code (but do not execute the code itself), and collect string variables that match a URI pattern. This investigation of the JavaScript parse tree allows us to also detect trackers that are dynamically added into the website's HTML from Javascript code, e.g., \texttt{scorecardresearch.com} in the following example:

\begin{verbatim}
<script type="text/javascript">document.write(
unescape("%3Cscript src='" + 
(document.location.protocol == "https:" ? 
"https://sb" : "http://b") + 
".scorecardresearch.com/beacon.js' 
%3E%3C/script%3E"));</script>
\end{verbatim}

\newpage
This approach allows for scaling to unprecedented data sizes, exceeding related work by more than an order of magnitude in the number of domains and pages analyzed. However, we note that it also imposes a set of limitations in comparison to data collection techniques that rely on instrumenting real browsers or on the collection of users' HTTP traces (cf.\ \citealp{Krishnamurthy2006, Roesner2012, Libert2015, Englehardt16}). One limitation of the use of the static CommonCrawl corpus is that transient trackers are excluded, i.e., trackers which are generated from JavaScript code that is fetched and executed dynamically at rendering time of the web page. Furthermore, we do not manually investigate the data recorded by each of the trackers, nor information that may influence tracking such as caching policies. We focus our analysis on the ability of third-parties to track (instead of the detailed tracking mechanisms) and on the high-level structural patterns of the tracking sphere.

We run a MapReduce \citep{Dean2010} implementation of our extractor in a massively parallel manner on the CommonCrawl 2012 web corpus, via the Amazon Elastic MapReduce service. In aggregate, the extraction takes about 60 hours using a cluster of 20 `m4.2xlarge' instances (8 virtual CPUs and 32~GB of RAM per instance). 

\subsection{Dataset Statistics and Characteristics}
 
Our dataset contains 3,536,611,510 web pages, which we aggregate by pay-level domain to create the \textit{bipartite third-party network}. This network represents the embeddings of 12,756,244 third-parties (the right vertex set) into 41,192,060 pay-level domains (the left vertex set). We create an edge between a website domain and a third-party domain whenever we find the third-party embedded at least once in the web pages belonging to a website.

\begin{table}
  \caption{Networks used in our study.}
  \label{tab:datasets}
  \centering
  \scalebox{0.9}{
    \begin{tabular}{r l l}
      \toprule
      \textbf{Count} & \textbf{Entities} & \textbf{Role}\\
      \midrule
      \multicolumn{3}{l}{\textit{Bipartite third-party network}}\\
      \midrule
      41,192,060 & Website PLDs & Left vertex set\\
      12,756,244 & Third-party PLDs & Right vertex set\\
      140,613,762 & Third-party embeddings & Edge set \\
      \midrule
      \multicolumn{3}{l}{\textit{Bipartite tracking network}}\\
      \midrule
      41,192,060 & Website PLDs & Left vertex set\\
      355 & Tracking PLDs & Right vertex set\\
      36,982,655 & Third-party embeddings & Edge set \\  
      \midrule      
      \multicolumn{3}{l}{\textit{Hyperlink network (obtained from webdata commons)}}\\
      \midrule
      41,192,060 & Website PLDs & vertex set\\
      623,056,313 & Hyperlinks & edge set \\    
      \bottomrule
    \end{tabular}
  }
\end{table}

Next, we order the third-parties by the fraction of domains on which they occur and select the top third-parties for manual labeling. We leave unlabeled the long tail of third-parties that are only present on a few thousand out of the 41 million domains to keep the labeling effort manageable. We enrich the data for the resulting 1,375 third-parties as follows: We determine the registration countries and registering organizations for the third-party domains. Then, we manually check the websites of the domains to determine their owning companies, and label the third-parties according to their purpose and business model. We find 355 pay-level domains to belong to tracking services, as these match our definition of trackers introduced in~Section~\ref{sec:foundations}. Thereby, we construct the \textit{bipartite tracking network}, which represents the 36,982,655 embeddings of the 355 tracking services in the 41,192,060 website pay-level domains of our corpus. Table~\ref{tab:datasets} gives an overview over the datasets we use for this study. We provide our obtained and curated data to the scientific community.\footnote{available at \url{https://ssc.io/trackingthetrackers/}}

Additionally, we obtain the network of 623,056,313 hyperlinks between the pay-level domains in CommonCrawl~\citep{WebdataCommonsHyperlinkGraph} for analyses that leverage the structure of the web.

\section{Global Analysis of Tracking}\label{chap:global} 
In this section, we study the extracted third-party and tracking networks to gain insights about the distribution of online tracking services on the Web. We want to answer the question to what proportion the browsing behavior of users on the Web is visible to particular tracking services. Furthermore, we are interested in how strongly the tracking capabilities differ among various services. Unfortunately, there is a lack of openly available data sources that allow to quantify the number of visitors over time for the 41 million extracted pay-level domains in 2012. Therefore, we resort to PageRank~\citep{Page1999}, a measure of the relevance of websites, as proxy for ranking them by the traffic which they attract. We compute the PageRank distribution of the domains in the hyperlink network to get an importance ranking for all the pay-level domains in our corpus. Next, we use this distribution to define our main ranking measure for the tracking capability of a third-party.

\subsection{Ranking Third-Parties}\label{sec:ranking}

We derive a ranking measure for third-parties from the PageRank value $p$ of the pay-level domains in the hyperlink network as follows. Let $D$ denote a set of domains to inspect, e.g., all pay-level domains belonging to a certain top-level domain, and let $t(D)$ denote the subset $\{d \,|\, d \in D \wedge t \text{ embedded in } d\}$ of domains contained in $D$ having third-party~$t$ embedded. We define the \textit{rank share} $\rho_{D,t}$ of a third-party $t$ in domain set $D$ as the sum of the PageRank values of domains from $D$ that have $t$ embedded, normalized by the overall sum of the PageRank values of domains in~$D$: 
\begin{equation}
         \rho_{D,t} = \frac{\sum_{i \in t(D)} p_i}{\sum_{i \in D} p_i}  \tag{rank share}
\end{equation}

In some cases, it is useful to look only at the number of pay-level domains visible to a third-party, regardless of their individual relevance.
In these cases, we additionally report the \textit{domain share} $d_{D,t}$ of third-party $t$ in domain set $D$ as the number of domains $t(D)$ in $D$ having $t$ embedded, normalized by the overall number of domains in~$D$: 
\begin{equation}
        d_{D,t} = \frac{\vert t(D) \vert}{\vert D \vert} \tag{domain share}
\end{equation}

\begin{figure}
  \centering    
  \includegraphics[scale=0.75]{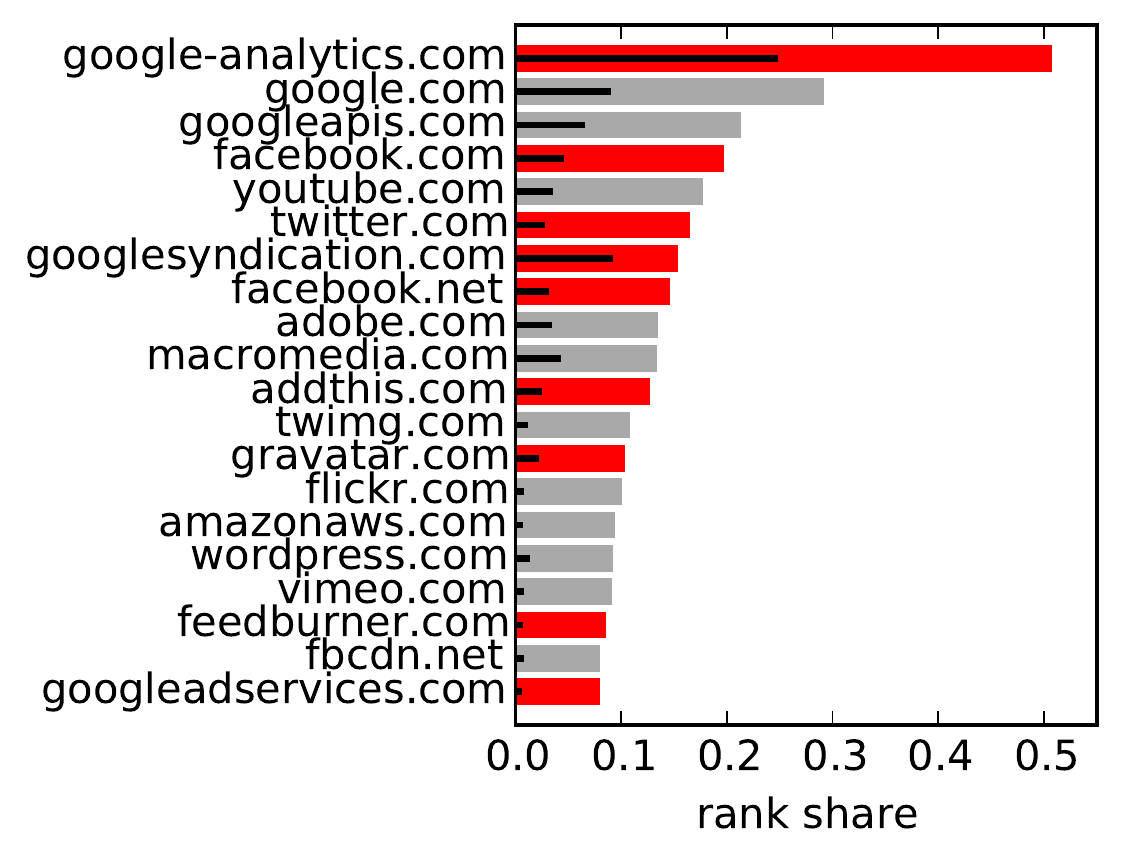}
  \caption{
    Google owns the three tracking websites with the largest share. 
    The twenty most common third-parties are shown by rank share. Tracking third-parties are highlighted in red, and domain share is shown as small black bars.
  }
  \label{fig:top-thirdparties-pagerank}
\end{figure}

\subsection{Predominant Third-Parties}

We first compute the third-parties with the highest rank share in our corpus. Figure~\ref{fig:top-thirdparties-pagerank} shows the top twenty third-parties by rank share (and also reports their domain share as small black bars). By far, the most common third-party is \texttt{google\allowbreak anal\allowbreak ytics.com} with a rank share of $0.507$, which implies that the pay-level domains embedding \texttt{google\allowbreak anal\allowbreak ytics.com} amount to more than half of the mass of the PageRank distribution in our corpus. Its domain share is $0.248$, implying that it is embedded on $24.8\%$ of all pay-level domains. We find that five out of the ten most dominant third-parties belong to Google. The next dominant family are social media-related third-parties such as \texttt{facebook.com}, \texttt{twitter.com} and \texttt{addthis.com}. 
On the lower end, we find content delivery services such as \texttt{twimg.com}, the image hosting platform of Twitter, the cloud platform Amazon Webservices (\texttt{amazonaws.com}) and Facebook's content delivery platform \texttt{fbcdn.net}. We highlight third-parties that match our definition of tracking services from Section~\ref{sec:foundations} in Figure~\ref{fig:top-thirdparties-pagerank}, which includes the dominant third-party \texttt{google\allowbreak anal\allowbreak ytics.com} as well as eight additional out of the twenty predominant third-parties. 

\subsection{Differences in Tracking Capability}
 
  \begin{figure}
    \centering
    \includegraphics[scale=0.3]{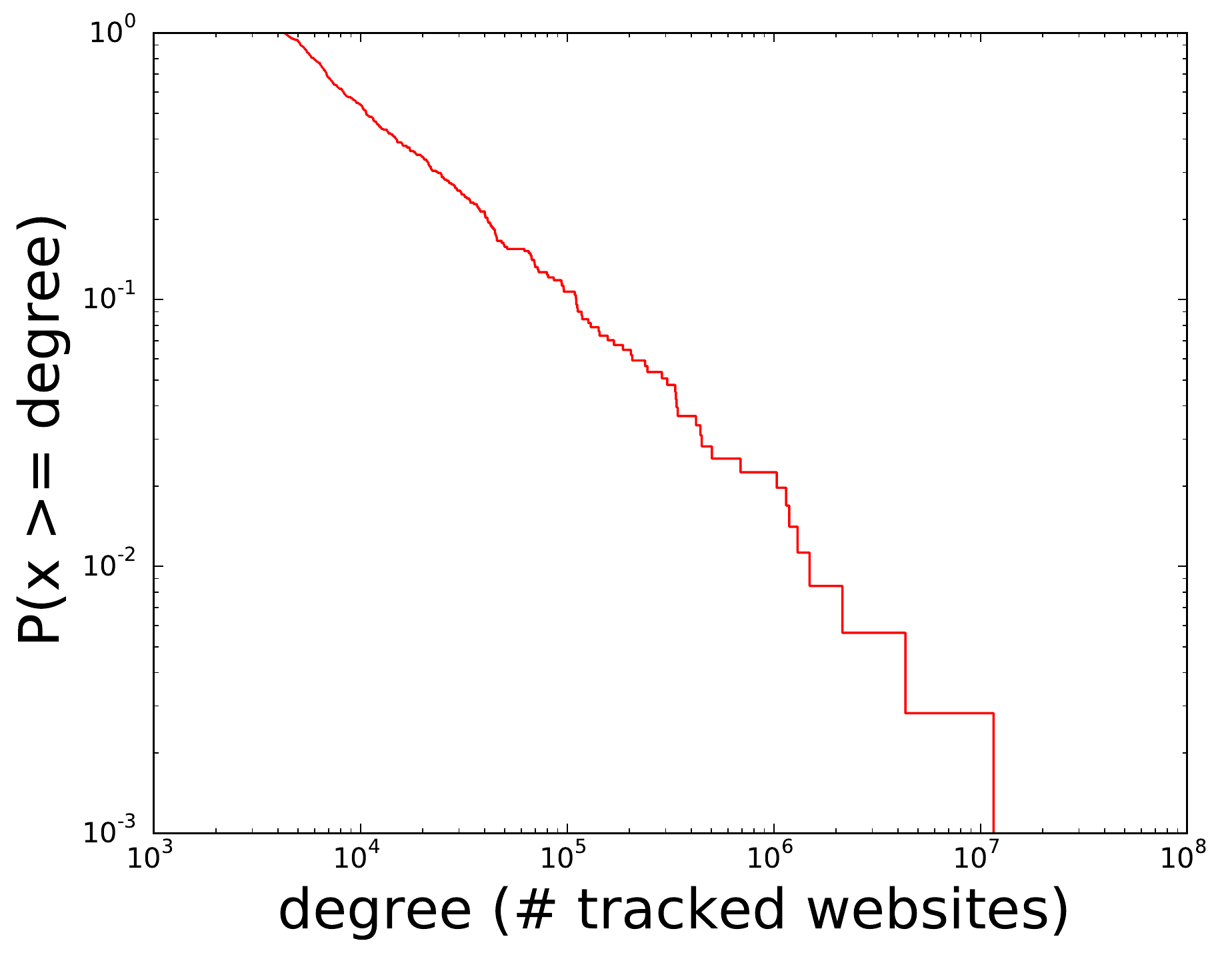}
    \caption{Cumulative distribution of the number of pay-level domains visible to tracking services in the bipartite tracking network.}
    \label{fig:degree-distribution}
  \end{figure}

Next, we study the differences in tracking capability between individual tracking services. To do that, we restrict the following analysis to the bipartite tracking network, which only contains the third-parties that match our definition of tracking services. We investigate the distribution of the number of pay-level domains visible to an individual tracking service. This distribution corresponds to the degree distribution of the right vertex set in the bipartite tracking network. Figure~\ref{fig:degree-distribution} shows the corresponding cumulative distribution function (on logarithmic axes) of this distribution. We encounter a highly disproportionate distribution: 50\% of the tracking services are embedded on less than ten thousand pay-level domains, while tracking services in the top 1\% of the distribution are integrated into more than a million pay-level domains. To assess whether this distribution follows a power law,
we apply the methodology by Clauset and colleagues \citeyearpar{Clauset2009}, which uses a goodness-of-fit test to find a minimum value of the degree from which a power law is valid, as well as estimates the power law exponent. 
We find that starting from degree 6,848, the distribution indeed follows a power law with exponent $1.725$, which is at the upper end of the range of exponents observed in many other hyperlink networks.\footnote{\url{http://konect.uni-koblenz.de/statistics/power}}

Additionally, we investigate the assortativity of the tracking network, and find that low degree domains (websites with few trackers) tend to connect to high-degree tracking services. This amounts to a dissortative mixing with a negative correlation coefficient of $-0.1863$.  
The found dissortativity is statistically significant to a $p$-value under 0.001. 
This encountered dissortativity is again consistent with findings for other hyperlink networks.\footnote{\url{http://konect.uni-koblenz.de/statistics/assortativity}}

  \begin{figure}
      \centering
      \includegraphics[scale=0.4]{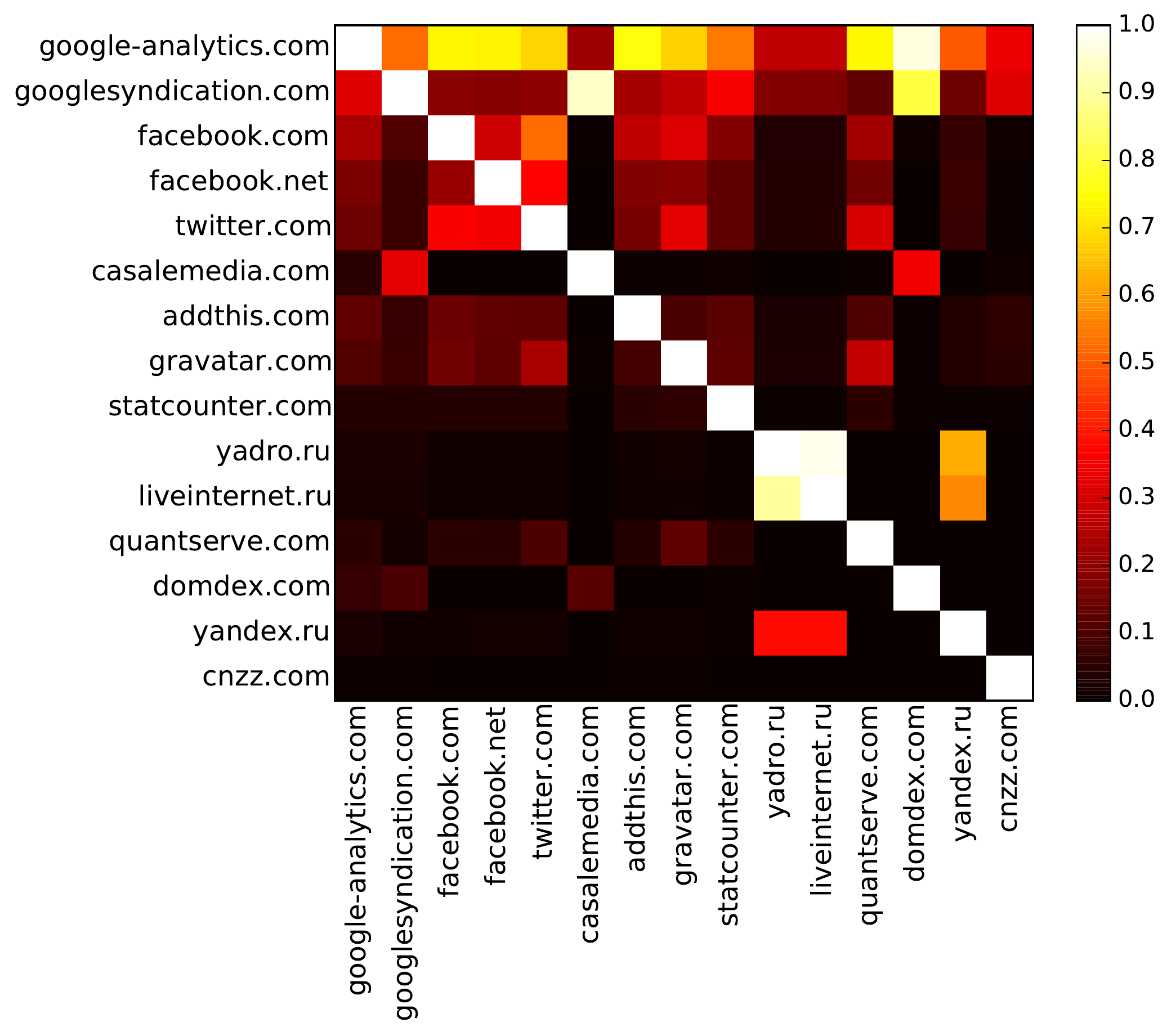}
      \caption{Heatmap representing the conditional probabilities $P(t_1 \mid t_2)$ of encountering tracking third-party $t_1$ (Y axis) on a website pay-level domain, given that this website has tracking third-party $t_2$ (X axis) embedded.}
      \label{fig:superposition}
  \end{figure}

   \begin{figure*}[th]
     \centering
     \includegraphics[width=0.90\textwidth]{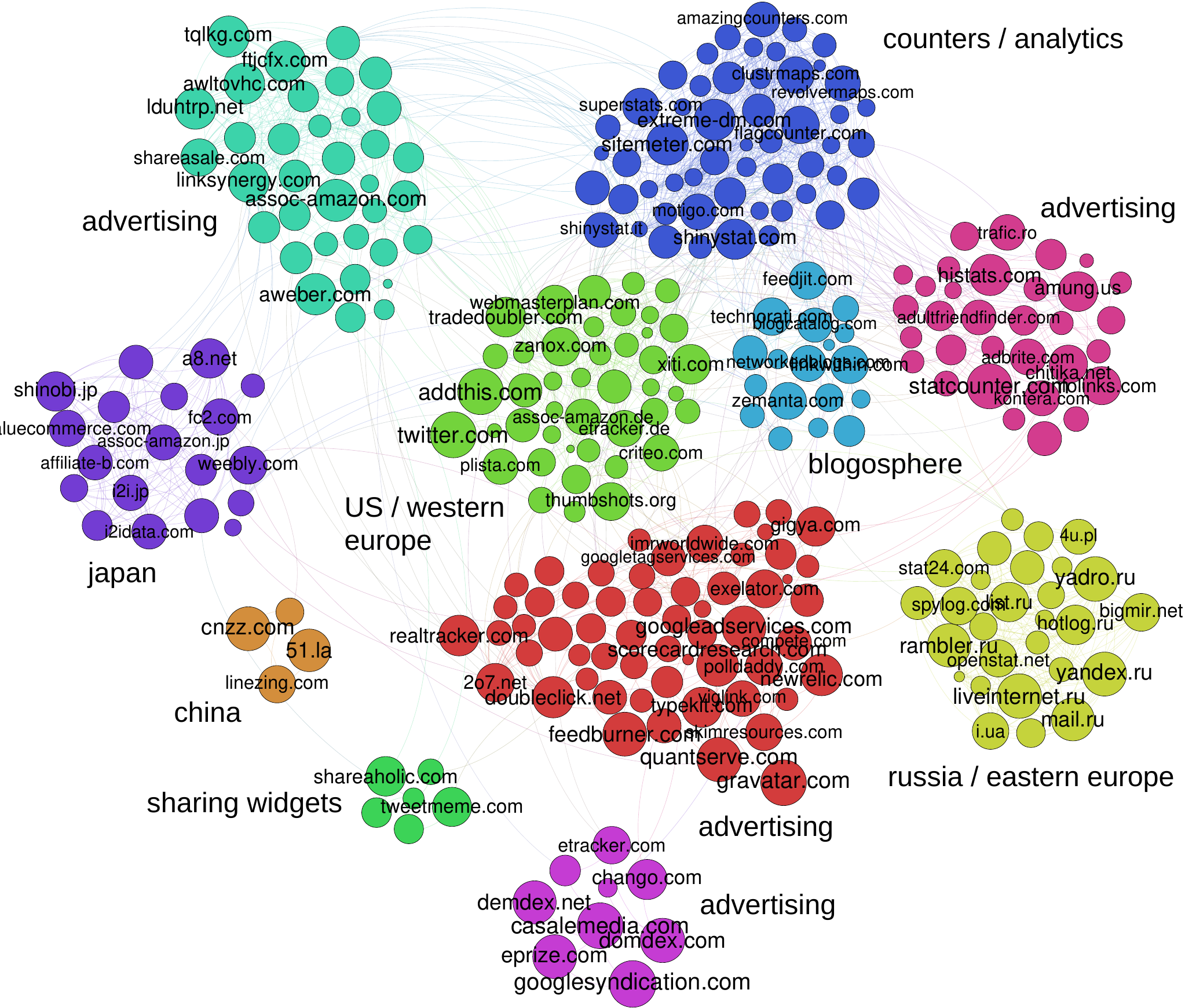}
     \caption{Cluster analysis of the co-occurrence network of tracking third-parties using modularity maximization. We manually derive the labels by investigating the distribution of third-party categories and registration countries inside the clusters.}
     \label{fig:clustering-cooccurrence}
   \end{figure*}

\subsection{Tracker Co-Occurrences}

We now focus on the relationships between individual pairs of tracking third-parties, in order to answer questions of the form \emph{What proportion of the website domains that embed \texttt{facebook\allowbreak .com} also embed \texttt{twitter.com}?} We therefore compute the conditional probability $P(t_1 \mid t_2)$ of encountering a tracker $t_1$ on a website pay-level domain, given that it already has another tracker $t_2$ embedded. 
Figure~\ref{fig:superposition} illustrates the results for all pairs within the fifteen most predominant tracking services using a heat map. From this experiment, we make multiple observations. First, we observe the predominant role of \texttt{google-\allowbreak analytics.com}, having by far the highest probability of co-occuring with any other service, consistent with its wide spread over the Web. Furthermore, we find a high agreement in the probability patterns for social media-related services such as \texttt{facebook.com}, \texttt{twitter.com} and \texttt{gravatar.com}. Additionally, the heatmap hints at country-specific patterns, hence the isolation of Russian tracking third-parties \texttt{yadro.ru}, \texttt{liveinternet\allowbreak .ru} and \texttt{yandex.ru}.

\begin{figure*}[t]
    \centering
    \includegraphics[width=1.0\textwidth]{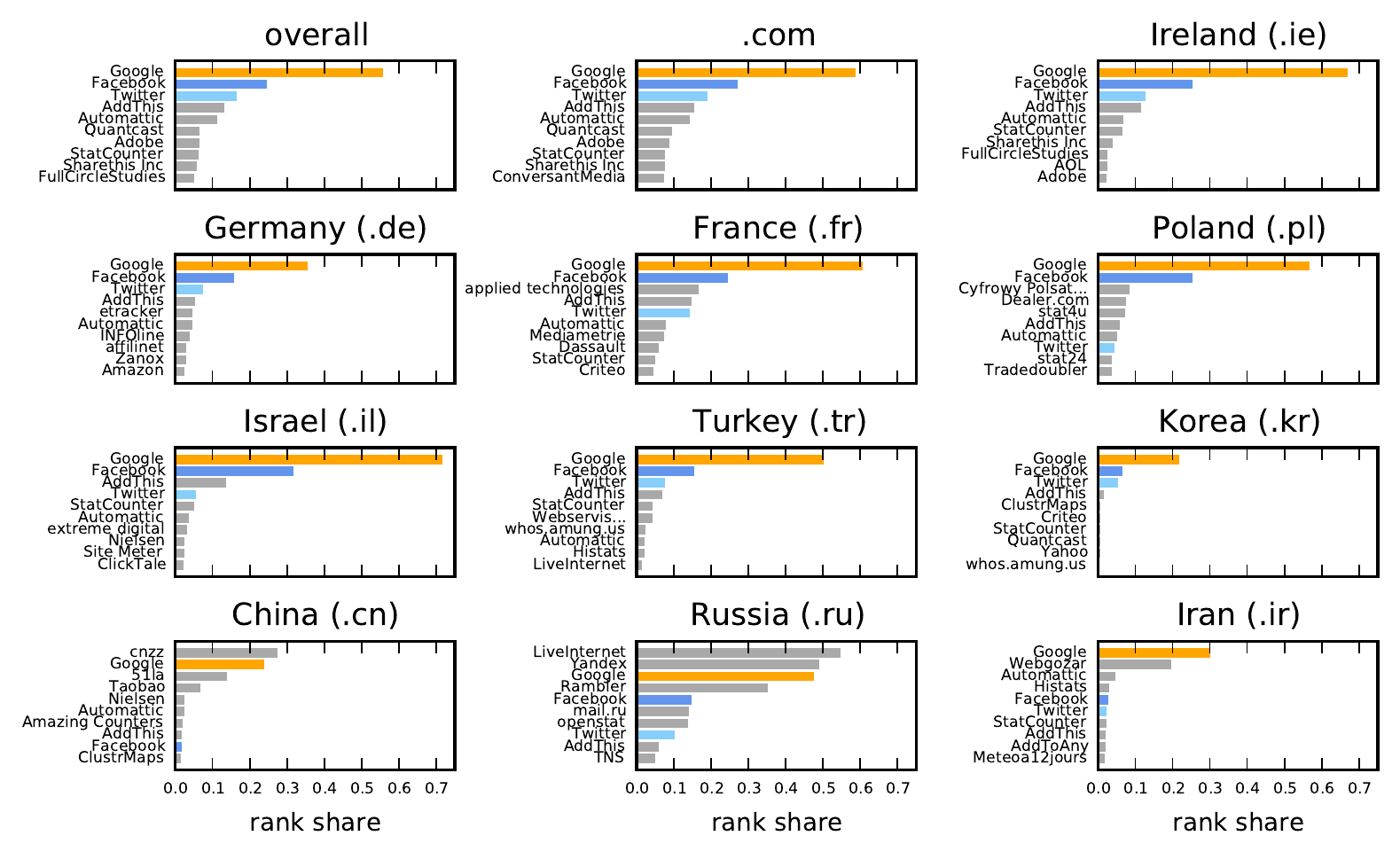}
    \caption{
        The ten companies per top-level domain with the highest rank share for a selection of countries. Google, Facebook and Twitter are highlighted in orange,  dark blue, and light blue respectively. 
    }
    \label{fig:top-companies}
\end{figure*}

\subsection{Cluster Analysis of Co-Occurrences}

We focus our analysis on the country-specific and media-specific patterns that the previous analysis pointed to in the data. We proceed in two steps to uncover these patterns: First, we compute the co-occurrence network of tracking services to find pairs of trackers that occur on the same sites very often. Second, we cluster this network using the Louvain method for community detection \citep{Blondel2008}, and inspect the resulting clusters.

We compute the adjacency matrix of the co-occurrence network as the one-mode projection $A = B^T B$, with $B$ being the bi-adjacency matrix of the tracking network \citep{kunegis:bipartivity}. We use log-likelihood ratio-based $G^2$ tests \citep{Dunning1993} to prune its edges; thereby we only retain tracker pairs which occur more often than expected. Finally, we cluster the resulting network via greedy modularity maximization with the Louvain method. We run a grid search on the hyperparameters of the clustering algorithm and choose the cluster assignments which result in the highest modularity $\frac{1}{2m} \sum_{ij}  [ A_{ij} - \frac{k_i k_j}{2m} ] \delta(c_i, c_j)$. Here, $m$ denotes the number of edges, $k_i$ as well as $k_j$ refer to the degrees of vertices $i$ and $j$, and $c_i$ and $c_j$ represent the cluster assignments of vertices $i$ and $j$. $A_{ij}$ is the weight of the edge connecting vertices $i$ and $j$, and $\delta(c_i, c_j)$ is the Kronecker delta which gives $1$ if $c_i = c_j$ and $0$ otherwise. 

As a clustering technique, modularity maximization has an intuitive interpretation as a cost function: For the edges between all vertices $i$ and $j$ assigned to the same cluster, it measures the difference between the observed edge weight $A_{ij}$ and the expected edge weight $\frac{k_i k_j}{2m}$ of a randomly rewired graph. Thereby, a high modularity corresponds to dense connections inside the clusters, and sparse connections outside. We clean up the clustering result by only retaining the 2-core of the resulting co-ocurrence graph, which has 329 trackers connected by 1,857 edges. 

Figure~\ref{fig:clustering-cooccurrence} illustrates the eleven clusters of trackers found in our dataset. In order to label the resulting clusters, we look at  types of tracking services (e.g., advertisers or social media plugins), registration countries of their domains, as well as at the top-level domains of the website domains which embed the trackers. We compare the distributions of these attributes in the clusters to the corresponding distributions of these attributes in the corpus as a whole. We encounter four clusters that clearly exhibit country-specific patterns. All third-parties to which we could attribute a country in the cluster containing \texttt{twitter.com} and \texttt{addthis.com} belong to the United States or to Western European countries (e.g., Germany, France, Great Britain, Sweden, Finland and Switzerland). This finding is supported by the fact that leading European advertising and affiliate marketing platforms such as \texttt{criteo.com} and \texttt{zanox.com} belong to this group. The cluster containing \texttt{liveinternet.ru} and \texttt{yadro.ru} is mostly comprised of third-parties from Eastern European countries (e.g., Russia, Poland and Belarus). Interestingly, it also contains 7\% of third-parties registered in the Netherlands. We attribute this to the fact that many Russian internet companies are legally based in the Netherlands for reasons of taxation \citep{RBTH2014}. Next, we encounter a cluster containing \texttt{shinobi.jp} which comprises mostly Japanese third-parties. Finally there is a small cluster which contains a majority of Chinese third-parties, such as \texttt{cnzz.com}. 

Not all clusters lend themselves to a geographic interpretation however. The remaining clusters exhibit similarities in the type of trackers embedded or in the type of website tracked: We find a very large cluster of mostly web counter and web analytics services, with representatives such as \texttt{shinystat.com} and \texttt{amazingcounters.com}. Additionally, we find a small cluster of sharing widgets (e.g., \texttt{shareaholic.com} and \texttt{tweetmeme.com}). A third cluster, represented by \texttt{technorati.com} and \texttt{networked\allowbreak blogs.com}, consists mostly of domains related to blogging. The remaining clusters mostly consist of advertising services.

\section{Country-Specific Analysis}\label{chap:country}
In the previous sections, we have found that the clustering of the co-occurrence network exhibits clear country-specific patterns. In order to better understand the relationships between individual countries and their distribution of web tracking, we conduct a set of deeper analyses on the country level. Every country is represented by a top-level domain such as \texttt{.de} for Germany and \texttt{.fr} for France. We omit the United States which in addition to its country code-based top-level domain \texttt{.us} also uses \texttt{.com}, which in turn is also widely used by non-US websites.  We look at the 50 predominant top-level domains in our corpus and their corresponding countries belonging to many continents, ranging from Germany (\texttt{.de}) hosting 2.87 million pay-level domains to Thailand (\texttt{.th}) with 15,858 pay-level domains.

\subsection{Predominant Tracking Companies\\ per Country}\label{sec:predominant}

We compute the rank share of tracking services in the subset of pay-level domains belonging to a specific country, represented by its top-level domain. We aggregate the results on company level in order to match a tracker to the country in which is owning company is based. Then, we compare the distribution of tracking companies in a specific country to the distribution of companies in the corpus as a whole, and to the distribution in the \texttt{.com} top-level domain. Figure~\ref{fig:top-companies} shows the results of this analysis for a selection of countries: Ireland, Germany, France, Poland, Israel, Turkey, Korea, China, Russia and Iran. We highlight the bars for the three globally most dominant companies: Google, Facebook and Twitter. These three companies have a special role, as we encounter them in the majority of top ten lists, in many cases accumulating the largest amount of rank share. Even among these three, Google has a preeminent position: we find it in a dominating role in the majority of countries, often with an amount of rank share that is more than double of what the second-placed company accumulates. We find that in many cases, the top ten companies consist of the three dominant US companies, Google, Facebook and Twitter, accompanied by a set of companies resident in the country under observation. Examples are Zanox (affiliate marketing) and INFOnline (digital audience measurement) in Germany, Criteo (advertising) in France, as well as Yandex and LiveInternet in Russia. These country-resident companies hardly ever appear in the top ten list of another country. The pattern of dominance of Google, followed by Facebook and Twitter is present in the overall corpus as well as in the vast majority of countries; however there are a few notable outliers, e.g., China and Russia where country-resident companies such as Yandex and CNZZ outrank Google.

\subsection{Correlation Analysis of the Dominance of Google, Facebook and Twitter}

We further investigate the country specifics of the three companies with a dominating role: Google, Facebook and Twitter. We therefore define a simple, dichotomous measure of dominance per country. We say that these three companies have a dominating role in a country if they accumulate more than half of the mass of the rank share distribution in the top ten companies. We compute this measure for the 50 countries under investigation in our corpus, and encounter the dominance pattern for a vast majority of 46 countries. Only four countries do not exhibit this pattern: China, Russia, Iran and Ukraine.

\begin{table}[t]
  \caption{Correlation of various country-level indicators with the tracking dominance of Google, Facebook and Twitter.}
  \label{tab:correlations}
  \centering
  \begin{tabular}{l l}
    \toprule
    \textbf{Variable} & \textbf{Correlation}\\
    \midrule
    Democracy index & $0.662^{***}$\\
    Freedom of the press & $0.612^{***}$\\
    \midrule
    Proportion of English speakers & $0.343^{*}$\\
    \midrule
    Online ad spending per capita & 0.333\\
    US trade volume & 0.167\\
    Online ad spending ratio & 0.062\\
    \bottomrule
    * $p < 0.05$, $\,$ ** $p < 0.01$, $\,$ *** $p < 0.001$     
  \end{tabular}
\end{table}

We would like to gain insights into factors which explain these findings. Intuitively, there could be influential factors of different character, e.g., political (e.g.,~states purposefully discriminate foreign companies to uphold control over media), socio-cultural (e.g., a language is dominant which is very different from English, making it harder for foreign companies to enter the market) or economical (e.g., certain countries might be more attractive to advertisers due to higher ad spending). We leverage a set of additional data sources for this analysis: an index describing the democratic character of a country~\citep{EconomistDemocracy12}, another index measuring the degree of freedom of the press in a particular country~\citep{FreedomHouse12}, data about the ratio of English speakers in a country's population~\citep{Wikipedia15}, as well as population counts~\citep{WorldBank15}. Furthermore, we include economic data about 
 US foreign trade~\citep{CensusGov15} and statistics about worldwide ad spending~\citep{CatchaDigital13}.

We compute several country-specific indicator variables from these additional datasets: Our political indicators consist of the \textit{democracy index} and the \textit{freedom of the press index}. For the latter, we simply revert the existing scale to make high values indicate high freedom of the press. We use the \textit{percentage of the population which speaks English} as socio-cultural indicator. Finally, we derive several economic indicators. We compute the \textit{online ad spending per capita} as the sum of digital and mobile ad spending per country normalized by its population. The \textit{online ad spending ratio} is the ratio of the sum of digital and mobile and spending to the overall media ad spending. Lastly, the \textit{US trade volume} is the sum of imports and exports of the US with the given country, normalized by the size of the population of the country. 

We calculate the point-biserial correlation coefficient of the indicators to our dichotomous dominance variable, in order to measure their strength of association. Table~\ref{tab:correlations} shows the results. We find a very strong and at the same time statistically significant correlation with the political indicators, democracy index and freedom of the press. The socio-cultural indicator, amount of English speakers, is only moderately correlated and only statistically significant at the 0.05 level. The economic indicators, online ad spending per capita, US trade volume and online ad spending ratio show low to moderate correlation, which is not statistically significant. These findings are surprising as they indicate that a positive characteristic such as freedom of the press is accompanied by a potentially very negative characteristic: the recording of people's browsing behavior by companies outside of the legal control of the countries institutions.

\section{Content-Specific Analysis}\label{chap:contents}

\begin{figure}
  \centering
  \includegraphics[scale=0.3]{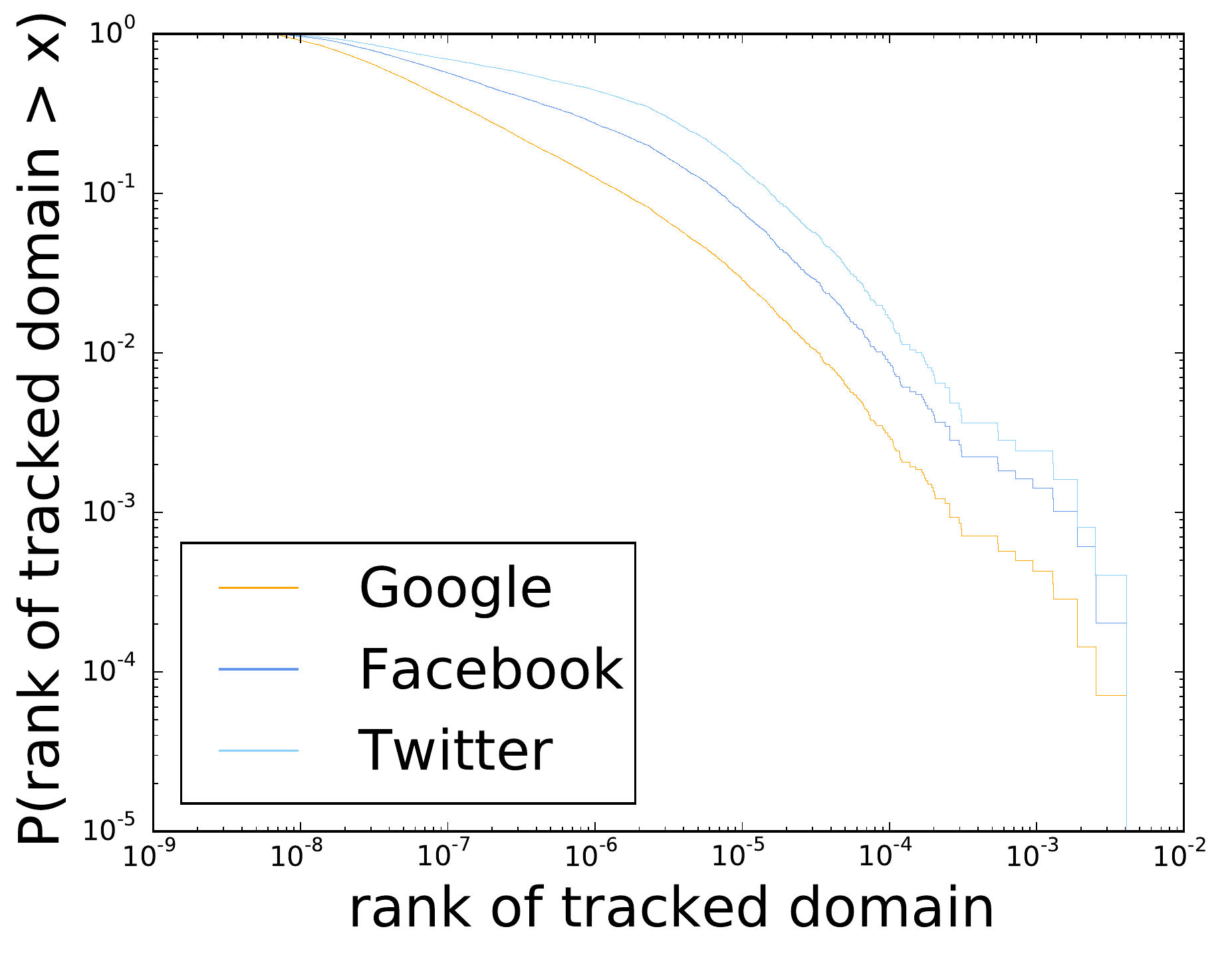}
  \caption{Per-company cumulative distribution of the PageRank of tracked domains with highly privacy-critical content.}
  \label{fig:privacy-cdf}
\end{figure}

In this section, we investigate how the presence and amount of trackers in a website is related to its content. We focus on comparing domains with highly privacy-critical, intimate content to domains with less critical content. We want to know whether there is a reduction in tracking due to the intimate nature of the content. We leverage the DMOZ database \citeyearpar{Dmoz}, a large, human-edited directory of the Web which provides an extensive labeling of websites. We define four highly privacy-critical categories of websites, as well as four less privacy-critical categories in order to compare the distribution of trackers among them. We associate every such defined category with a set of DMOZ labels and extract the domains of all websites with that label. We repeat this for all available translations of the label in up to 41 different languages. Next, we intersect these domains with the domains in our corpus. We use the resulting set of labeled domains as a representative for the category.

This process results in 24,026 \textit{health}-related domains, originating from DMOZ labels such as \texttt{Health/\allowbreak Mental\_\allowbreak Health} and \texttt{Health/\allowbreak Medicine/\allowbreak Surgery} and 2,436 \textit{addiction}-related domains, which have labels such as \texttt{Health/\allowbreak Addictions}. Furthermore we find 3,421 \textit{sexuality}-related domains, from labels such as \texttt{Society/\allowbreak Gay,\_\allowbreak Lesbian,\_\allowbreak and\_\allowbreak Bisexual} and 611 domains related to \textit{gender identity}, which we get by looking at labels such as \texttt{Society/\allowbreak Transgendered}. We chose these categories, as the association of a person with the topics discussed in these websites could potentially result in discrimination against the person (e.g.,~less chances during job search due to an illness), or even physical harm (e.g.,~prosecution of homosexual persons with prison or even the death penalty in some countries). As less privacy-critical categories, we choose \textit{cooking \& food} (40,698 domains), \textit{soccer} (11,677 domains), \textit{television} (5,793 domains) and \textit{video games} (12,016 domains). Overall, this analysis includes 91,813 distinct labeled domains.

\subsection{Overall Avoidance of Trackers on Highly Privacy-Critical Websites}
We perform a statistical test to find out whether in general, privacy-critical websites are more likely to contain trackers than privacy-noncritical websites.  Table~\ref{tab:privacy-overall} summarises the total number of analysed websites in both categories with and without trackers.
We observe that across all types of trackers combined, privacy-critical websites are less likely to contain trackers than privacy-noncritical websites.  Indeed, about 90\% of websites with less privacy-critical content contain trackers, while only about 60\% of websites with highly privacy-critical content do. 
This result is significant to a $p$-value of $p < 0.001$. 
While this confirms that the choice of including trackers on a website seems to correlate with the topic of the website, indicating that trackers are avoided on possibly privacy-incriminating topics, the numbers still show that a majority of websites with highly privacy-critical content \emph{do} contain trackers. 
This result also raises the question of \emph{which} trackers are employed on which type of website, and whether some type of trackers are preponderant on any type of website, highly privacy-critical or less so.  Therefore, the next experiment investigates each tracker separately. 

\begin{table}
  \caption{
    Number of highly privacy-critical and less privacy-critical websites with and without trackers.
    Overall, less privacy-critical websites are more likely to contain trackers than highly privacy-critical websites. 
    \label{tab:privacy-overall}
  }
  \centering
  \scalebox{0.9}{
  \begin{tabular}{ l r r }
    \toprule
    & \textbf{Highly critical} & \textbf{Less critical} \\
    \midrule
    \textbf{Without trackers} & 11,001 & 7,014 \\
    \textbf{With trackers} & 16,697 & 57,850 \\
    \midrule
    \textbf{Fraction with trackers} & 60.3\% & 89.2\% \\
    \bottomrule
  \end{tabular}
  }
\end{table}

\subsection{Trackers Specifically Present on\\ Highly Privacy-Critical Websites}

\begin{figure}
  \centering
  \includegraphics[scale=0.6]{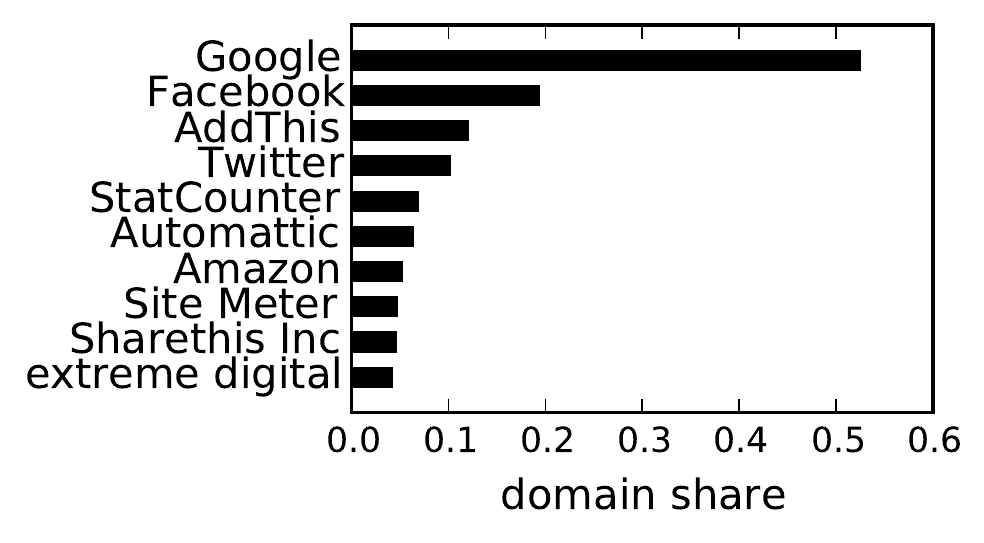}
  \includegraphics[scale=0.6]{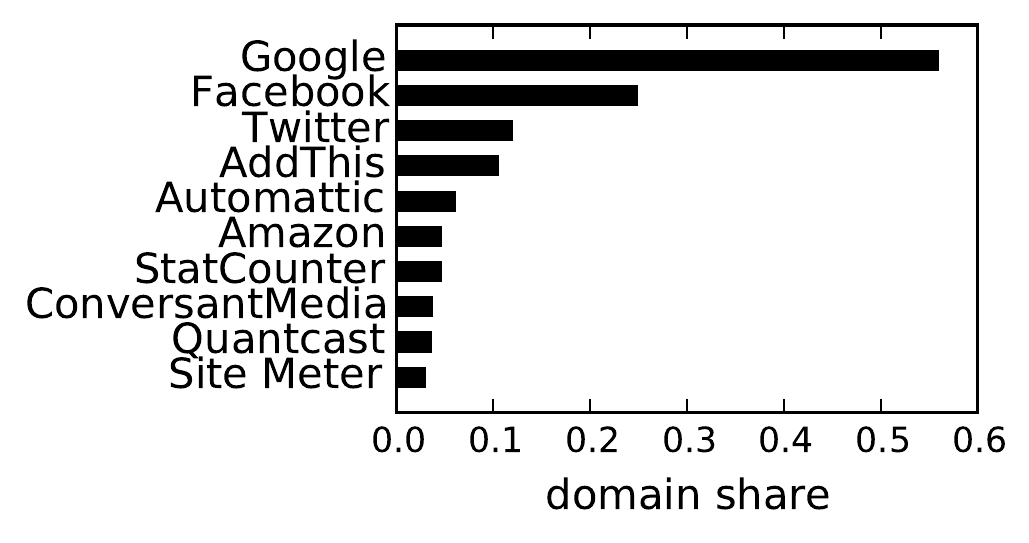}
  \caption{
    Top ten companies embedded on highly privacy-critical (top) and less privacy-critical domains (bottom).
  }
  \label{fig:domainshareboth}
\end{figure}

In this experiment, we want to find out for each tracker whether it is represented more on privacy-critical or less-critical sites.  We perform a statistical test based on the null hypothesis that the trackers are distributed randomly among all websites, and compare each tracker's distribution to that.  The results are shown in Table~\ref{tab:privacy-by-tracker}. We observe that the big three trackers (Google, Facebook, and Twitter) are less prevalent on highly privacy-critical websites.  This is consistent with the overall results that websites containing highly privacy-critical content avoid trackers.  However, other trackers are more prevalent on highly privacy-critical websites than on less privacy-critical ones.  These include AddThis, StatCounter, Amazon, Sharethis, Site~Meter, Adobe, and many more.  These trackers are from smaller companies than the big three, but all have in common that they are likely to be perceived not as \emph{tracking} services, but as added functionalities on a website, such as visitor counters, and sharing buttons, etc.  We conjecture that the fact that these trackers are perceive as less of a threat to privacy leads them to not be avoided by hosters of highly privacy-critical websites.  Nonetheless, it remains the case that even among websites with highly privacy-critical content, the most common trackers are Google, Facebook, AddThis and Twitter as shown in Figure~\ref{fig:domainshareboth}, i.e., the big three overall trackers are present in the top four spots, with only AddThis have a larger share of websites than Twitter.  Thus, the next experiments analyses the top three trackers specifically. 

\begin{table*}
  \caption{
    Classification of trackers by the type of website (highly privacy-critical or less privacy-critical) they are more prevalent on.  The table contains all trackers among the 20 most used trackers (by number of websites) which are significantly present for each type of website. 
    \label{tab:privacy-by-tracker}
  }
  \centering
  \begin{tabular}{ p{8.5cm} p{8.5cm} }
    \toprule
    \textbf{Prevalent on highly privacy-critical sites} & \textbf{Prevalent on less privacy-critical sites} \\
    \midrule
    AddThis\textsuperscript{***}
StatCounter\textsuperscript{***}
Amazon\textsuperscript{***}
Sharethis Inc\textsuperscript{***}
Site Meter\textsuperscript{***}
Adobe\textsuperscript{***}
extreme digital\textsuperscript{***}
AddToAny\textsuperscript{***}
Disqus\textsuperscript{**}
ClustrMaps\textsuperscript{***}
ShinyStat\textsuperscript{***}
Yahoo\textsuperscript{*}
Boardhost\textsuperscript{***}
applied technologies\textsuperscript{**}
CommissionJunction\textsuperscript{*}
Microsoft\textsuperscript{***}
Histats\textsuperscript{*}
Technorati\textsuperscript{***}
Motigo\textsuperscript{***}
fav.or.it.\textsuperscript{*}

    &
    Google\textsuperscript{***}
Facebook\textsuperscript{***}
Twitter\textsuperscript{***}
LiveInternet\textsuperscript{***}
FullCircleStudies\textsuperscript{***}
Rakuten\textsuperscript{*}
Yandex\textsuperscript{***}
Rambler\textsuperscript{***}
Nielsen\textsuperscript{***}
AOL\textsuperscript{***}
Tradedoubler\textsuperscript{***}
FC2\textsuperscript{***}
Samurai Factory\textsuperscript{***}
mail.ru\textsuperscript{***}
New Relic\textsuperscript{***}
A8\textsuperscript{***}
Chartbeat\textsuperscript{***}
ValueCommerce\textsuperscript{***}
VibrantMedia\textsuperscript{***}
INFOline\textsuperscript{***}

    \\
    \bottomrule
    * $p < 0.05$, $\,$ ** $p < 0.01$, $\,$ *** $p < 0.001$ 
  \end{tabular}
\end{table*}

\subsection{Distribution of Top Trackers in\\ Privacy-Critical Websites}
We compute the domain share and rank share of the three top trackers Google, Facebook and Twitter, as determined in Section~\ref{sec:predominant}, for all the domains in a particular category, and list the results in Table~\ref{tab:privacy-share}. Google has by far the highest domain-share in all categories, indicating that it is present on exceedingly more domains than any other company (cf.\ Table~\ref{tab:privacy-domainshare}). On average, we encounter Google on 51\% of domains per category, in contrast to Facebook and Twitter, which we see on 26\% and 14\% of domains only. Surprisingly, the picture changes once we look at rank share instead of domain share, as shown in Table~\ref{tab:privacy-rankshare}: Google has less rank share than Facebook. That suggests that Facebook tracks less but more prominent domains (in terms of PageRank) for our defined categories. To further investigate this finding, we plot the cumulative distribution of ranks corresponding to tracked domains per company of the highly privacy-critical categories in log-log scale in Figure~\ref{fig:privacy-cdf}. Investigating this distribution confirms our hypothesis that on average, Google occurs on lower-ranking domains than Facebook and Twitter. 

\begin{table*}[!htb]
    \caption{Tracking capabilities of Google, Facebook and Twitter on different categories of domains.}
    \begin{subtable}{.45\linewidth}
      \centering
        \caption{Domain share on highly privacy-critical categories and less privacy-critical categories of domains.}
        \begin{tabular}{l r r r }
          \toprule
           & Google & Facebook & Twitter\\
          \midrule
          Health & \textbf{0.473} & 0.157 & 0.081 \\
          Addiction & \textbf{0.427} & 0.152 & 0.092 \\
          Sexuality & \textbf{0.521} & 0.299 & 0.158 \\
          Gender Identity & \textbf{0.462} & 0.295 & 0.201 \\
          \midrule
          Cooking & \textbf{0.482} & 0.176 & 0.070 \\
          Soccer & \textbf{0.519} & 0.301 & 0.160 \\
          Television & \textbf{0.617} & 0.394 & 0.239 \\
          Video Games & \textbf{0.590} & 0.268 & 0.144 \\
          \bottomrule
        \end{tabular}
        \label{tab:privacy-domainshare}
    \end{subtable}%
    \hspace{\fill}
    \begin{subtable}{.45\linewidth}
      \centering
        \caption{Rank share on highly privacy-critical categories and less privacy-critical categories of domains..}
        \begin{tabular}{l r r r}
          \toprule
           & Google & Facebook & Twitter \\
          \midrule
          Health & 0.903 & \textbf{0.912} & 0.855 \\
          Addiction & 0.798 & \textbf{0.959} & 0.953 \\
          Sexuality & 0.888 & \textbf{0.989} & 0.834 \\
          Gender Identity & 0.802 & \textbf{0.990} & 0.914 \\
          \midrule
          Cooking & 0.890 & \textbf{0.925} & 0.788 \\
          Soccer & 0.904 & \textbf{0.981} & 0.840 \\
          Television & 0.909 & \textbf{0.976} & 0.839 \\
          Video Games & 0.916 & \textbf{0.965} & 0.834 \\
          \bottomrule
        \end{tabular}
        \label{tab:privacy-rankshare}
    \end{subtable} 
    \label{tab:privacy-share}
\end{table*}

\section{Related Work}
\subsection{Global Analysis of Web Tracking}

The analysis of the web tracking phenomenon has received much attention in the research community. The main novelty in this approach as compared to previous studies is that our data acquisition technique allows us to look at an order of magnitude more domains, and several orders of magnitude more pages. Furthermore, with the exception of work by \citet{Englehardt16}, no ranking is applied to tracker occurrences on domains, as our dataset allows to leverage the structure of the web graph to compute such a ranking (cf.\ Section~\ref{sec:ranking}).

\citet{Krishnamurthy2006} investigate web tracking and its privacy implications on about 1,000 domains by looking at the connections between visited domains and `hidden domains' (similar to our bipartite third-party network). They also conduct a country-specific analysis by looking at tracker on the top 100 domains for 68 countries, and find Google Analytics to be the dominating tracker. They extend this work in \citep{Krishnamurthy2009} where they investigate how the amount of tracking develops over time, both by single trackers as well as on a company level. They cover 1,200 English-language domains from Alexa top sites over four years between 2005 and 2008, and detect an increasing aggregation of private data by a decreasing number of trackers: the penetration of the top 10 trackers among popular sites has grown from 40\% to 70\% in the period of investigation. \citet{Roesner2012} develop a client-side method for detecting web trackers on 500 popular domains. They create a taxonomy of five different tracker types, based on how these manipulate browser state. Analogous to \citep{Krishnamurthy2006}, their data suggests a dominating role of Google Analytics. Furthermore, they estimate how much  of users web search browsing sessions is visible to trackers by sampling from query logs, and find that more than 20\% can be covered by several trackers. Finally, they develop a browser add-on called `ShareMeNot' as a defense mechanism, which removes cookies from certain web requests. In recent years, research has started to investigate tracking on a larger numbers of domains. \citet{Libert2015exposing} presents a quantitative analysis of tracking mechanisms on the top 1 million sites from the Alexa ranking. He finds that nearly nine in ten websites leak user data to third-parties, and that a handful of American companies (including Google, Facebook and Twitter) receive the vast bulk of this user data. Again, his data confirms the outstanding role of Google Analytics. Additionally, \citet{Libert2015} researches the privacy risks imposed by visiting health-related web pages on the web, based on the top 50 search results for about 2,000 common diseases. He finds that more than two thirds of these pages leak information about specific conditions, treatments and diseases to third-parties. \citet{Yu2016} develop a novel defense approach based on $k$-anomity. For that, they first process tracking data on 21 million pages from 350,000 domains, collected from traces of 200,000 internet users. Next, they show to leverage this data collectively and dynamically identify unsafe elements, and removed these elements from the requests. \citet{Englehardt16} present `OpenWPM', a comprehensive and scalable tracking measurement platform, which simulates real browsing. They use this platform to collect data on the top 1 million domains from Alexa. Analogous to our findings, they encouter a heavy-tailed distribution of tracking capability, postulate the dominating role of Google Analytics and state that many of the top-occurring third-party domains belong to Google.  This constitutes the only related study that also applies a weighting of occurrences of trackers on domains. They continue to investigate a wide variety of aspects of tracking, such as the low adoption of HTTPS encryption by trackers, they evaluate tracking protection techniques, and provide new insights into sophisticated tracking mechanisms, such as cookie syncing, `promiscuous' cookies, and previously unknown fingerprinting techniques. 

\subsection{Web Tracking by Social Networks}

Special focus has been given to certain actors that know about the real identities of internet users, such as online social networks. \citet{Krishnamurthy2009OSN} investigate the leakage of personally identifiable information from social networks to third-parties and suggest protection mechanisms. They extend this research to leakage from online social networks in \citep{Krishnamurthy2010}, and find that similar leakage happens there. \citet{Roosendaal2011} exemplarily investigates the privacy implications of the Facebook like-button. \citet{Chaabane2012} look at the tracking capabilities of the three major social networks (Facebook, Twitter and Google+) on the top ten thousand domains. Analogous to us, they find that tracking is encountered on all categories of sites, independent of their content. Furthermore, they show (based on browsing traces) that up to 77\% of a user's web profile can be reconstructed by these actors. 

\subsection{Tracking Mechanisms and Detection}

\citet{Eckersley2010} investigates `fingerprinting' techniques, based on version and configuration settings of modern web browers. They find that these techniques work surprisingly well at identifying individual users: for a random browser, only one in about three hundred thousand browsers will share its fingerprint. Furthermore, they developed a well-known test site\footnote{\url{https://panopticlick.eff.org}} for investigating such fingerprints. \citet{Acar2014} present an in-depth study of advanced tracking mechanisms such as browser fingerprinting via canvas images, re-spawning of HTTP cookies via the Adobe Flash plugin and cookie syncing between different trackers. They conclude that modern browers -- with the exception of the Tor Browser (\cite{Perry2013}) -- fail at effectively protecting users from most of these techniques. \citet{Bau2013} develop a machine learning-based approach for real-time tracker identification, which draws its features off a network derived from of script loading relationships in web pages. \citet{Kalavri2016} inspect a graph similar to our bipartite third-party network, which they derived from user traces collected by a web proxy. Their aim is to automatically identify trackers among the third-parties, based on structural properties of their collected graph. They show that simple classifiers such as a nearest neighbor approach, as well as label propagation techniques perform suprisingly well on this identification task. \citet{Englehardt2015} investigate the surveillance implications of passive eavesdroppers (e.g.,~intelligence agencies) piggybacking on existing tracking identifiers, and find that this allows them to reconstruct about two thirds of people's browsing~histories.

\subsection{Structural Analysis of CommonCrawl}

\citet{Spiegler2013} presents an exploratory analysis of the CommonCrawl 2012 corpus, computing elementary statistics such as the distribution of top-level domains, character encodings and media types. \citet{Lehmberg2014,Meusel2014,Meusel2015} study the fundamental graph structure of the web using the CommonCrawl 2012 dataset. Their findings suggest that the previously reported `bow-tie' structure from \citep{Broder2000} is an artifact of crawling process, and not a structural property of the web. Furthermore, their data shows that the distributions of in-degree, out-degree and sizes of strongly connected components of the page and host graph do not follow power laws.

\section{Conclusion and Discussion}
The scope of our analysis allows us to make several novel observations about online tracking. We found that 9 out of the 20 predominant third-party domains belong to trackers, and confirmed the extraordinary tracking capability of Google Analytics (cf.~\citealp{Roesner2012,Krishnamurthy2006}). Furthermore, we found that the distribution of the number of website domains tracked follows a power law, and that the overall tracking network is of a dissortative character. While there are many small trackers which are country-specific (e.g., to Germany and Japan), this is not true for the largest tracking services. These are Google, Facebook and Twitter, all US companies acting on a global scale, and representing the largest trackers in almost all countries.  The exception to this pattern are a small number of countries such as China, Russia and Iran, which all have little political ties to the US, and which represent outliers in terms of political factors such as democracy and freedom of the press. In particular, we could not determine a statistically significant correlation with economic factors such as amount of foreign trade between a country and the US, or with indicators related to ad spending. These findings lead us to the conclusion that the choice of tracking software made by website owners is largely independent of the website's topic, and mostly depends on political factors, mainly whether the country in question has a functional political relationship with the US. In economical terms, this confirms that social media companies in US-friendly countries such as Germany and Japan have a hard time getting large market share due to the dominance of US companies Google, Facebook and Twitter, while social media companies in countries such as Russia and China have better prospects due to the (voluntary or legislated) avoidance of US companies in those countries. Additionaly, our findings confirm that Google still operates tracking services on Chinese websites, despite its proclaimed retreat from the Chinese market \citep{TheGuardianGoogleChina}.

Our results indicate that the fact that a website covers highly privacy-critical topics does not imply the lack of tracking.  Even though the rate of tracked websites among those with highly privacy-critical content is lower than for other websites (60\% versus 90\%), the majority of such websites does still contain trackers.  One aspect of this high number is the apparent high number of trackers which are seemingly not perceived to be as dangerous as Google, Facebook and Twitter, namely those trackers with implement services benefitting the website itself, such as visitor counters.  For such websites, our results indicate that they are even more prevalent on privacy-critical websites. For end users, we can conclude that tracking is to be expected on all types of websites, regardless of the topic.  

\bibliographystyle{abbrv}
\bibliography{arxiv}  

\begin{thebibliography}{10}

\bibitem{WebdataCommonsHyperlinkGraph}
Web data commons -- hyperlink graphs.
\newblock \url{http://webdatacommons.org/hyperlinkgraph/}.

\bibitem{Disconnect}
Disconnect.
\newblock \url{https://disconnect.me/}, 2016.

\bibitem{Ghostery}
Ghostery.
\newblock \url{https://www.ghostery.com/}, 2016.

\bibitem{PrivacyBadger}
Privacy badger.
\newblock \url{https://www.eff.org/de/node/73969}, 2016.

\bibitem{Acar2014}
G.~Acar, C.~Eubank, S.~Englehardt, M.~Juarez, A.~Narayanan, and C.~Diaz.
\newblock The web never forgets: Persistent tracking mechanisms in the wild.
\newblock In {\em ACM CCS}, pages 674--689, 2014.

\bibitem{Bau2013}
J.~Bau, J.~Mayer, H.~Paskov, and J.~C. Mitchell.
\newblock A promising direction for web tracking countermeasures.
\newblock In {\em Proc. Web 2.0 Security and Privacy}, 2013.

\bibitem{Blondel2008}
V.~D. Blondel, J.-L. Guillaume, R.~Lambiotte, and E.~Lefebvre.
\newblock Fast unfolding of communities in large networks.
\newblock {\em Journal of Statistical Mechanics: Theory and Experiment},
  2008(10):P10008, 2008.

\bibitem{Broder2000}
A.~Broder, R.~Kumar, F.~Maghoul, P.~Raghavan, S.~Rajagopalan, R.~Stata,
  A.~Tomkins, and J.~Wiener.
\newblock Graph structure in the web.
\newblock {\em Computer networks}, 33(1):309--320, 2000.

\bibitem{CatchaDigital13}
{CatchaDigital}.
\newblock Worldwide ad spending forecast: Emerging markets, mobile provide
  opportunities for growth.
\newblock 2013.

\bibitem{Chaabane2012}
A.~Chaabane, M.~A. Kaafar, and R.~Boreli.
\newblock Big friend is watching you: Analyzing online social networks tracking
  capabilities.
\newblock In {\em 2012 ACM Workshop on Online Social Networks}, pages 7--12,
  2012.

\bibitem{Clauset2009}
A.~Clauset, C.~R. Shalizi, and M.~E. Newman.
\newblock Power-law distributions in empirical data.
\newblock {\em SIAM}, 51(4):661--703, 2009.

\bibitem{Dean2010}
J.~Dean and S.~Ghemawat.
\newblock Mapreduce: A flexible data processing tool.
\newblock {\em Communications of the ACM}, 53(1):72--77, 2010.

\bibitem{Dmoz}
{DMOZ}.
\newblock Directory of the web.
\newblock \url{https://www.dmoz.org/}, Jun 2016.

\bibitem{Dunning1993}
T.~Dunning.
\newblock Accurate methods for the statistics of surprise and coincidence.
\newblock {\em Comput. Linguist.}, 19(1):61--74, 1993.

\bibitem{Eckersley2010}
P.~Eckersley.
\newblock How unique is your web browser?
\newblock In {\em Privacy Enhancing Technologies}, pages 1--18. Springer, 2010.

\bibitem{EFFHealth}
{Electronic Frontier Foundation}.
\newblock Healthcare.gov sends personal data to dozens of tracking websites.
\newblock 2015.

\bibitem{Englehardt16}
S.~Englehardt and A.~Narayanan.
\newblock Online tracking: A 1-million-site measurement and analysis.
\newblock May 2016.

\bibitem{Englehardt2015}
S.~Englehardt, D.~Reisman, C.~Eubank, P.~Zimmerman, J.~Mayer, A.~Narayanan, and
  E.~W. Felten.
\newblock Cookies that give you away: The surveillance implications of web
  tracking.
\newblock In {\em WWW}, pages 289--299, 2015.

\bibitem{FreedomHouse12}
{Freedom House}.
\newblock Freedom of the press.
\newblock 2012.

\bibitem{Kalavri2016}
V.~Kalavri, J.~Blackburn, M.~Varvello, and K.~Papaginannaki.
\newblock Like a pack of wolves: Community structure of web trackers.
\newblock In {\em PAMS}, 2016.

\bibitem{Krishnamurthy2006}
B.~Krishnamurthy and C.~E. Wills.
\newblock Generating a privacy footprint on the internet.
\newblock In {\em ACM SIGCOMM}, pages 65--70, 2006.

\bibitem{Krishnamurthy2009OSN}
B.~Krishnamurthy and C.~E. Wills.
\newblock On the leakage of personally identifiable information via online
  social networks.
\newblock In {\em ACM workshop on Online Social Networks}, pages 7--12, 2009.

\bibitem{Krishnamurthy2009}
B.~Krishnamurthy and C.~E. Wills.
\newblock Privacy diffusion on the web: A longitudinal perspective.
\newblock In {\em WWW}, pages 541--550, 2009.

\bibitem{Krishnamurthy2010}
B.~Krishnamurthy and C.~E. Wills.
\newblock Privacy leakage in mobile online social networks.
\newblock In {\em USENIX Conference on Online social networks}, pages 4--4,
  2010.

\bibitem{Kunegis2013}
J.~Kunegis.
\newblock Konect: The koblenz network collection.
\newblock In {\em Proceedings of the 22nd International Conference on World
  Wide Web Companion}, pages 1343--1350. International World Wide Web
  Conferences Steering Committee, 2013.

\bibitem{kunegis:bipartivity}
J.~Kunegis.
\newblock Exploiting the structure of bipartite graphs for algebraic and
  spectral graph theory applications.
\newblock {\em Internet Math.}, 11(3):201--321, 2015.

\bibitem{Lehmberg2014}
O.~Lehmberg, R.~Meusel, and C.~Bizer.
\newblock Graph structure in the web: Aggregated by pay-level domain.
\newblock In {\em ACM Web Science}, pages 119--128, 2014.

\bibitem{Libert2015exposing}
T.~Libert.
\newblock Exposing the hidden web: An analysis of third-party http requests on
  1 million websites.
\newblock {\em International Journal of Communication}, 2015.

\bibitem{Libert2015}
T.~Libert.
\newblock Privacy implications of health information seeking on the web.
\newblock {\em Communications of the ACM}, 58(3):68--77, 2015.

\bibitem{Mayer2012}
J.~R. Mayer and J.~C. Mitchell.
\newblock Third-party web tracking: Policy and technology.
\newblock In {\em Security and Privacy (SP), 2012 IEEE Symposium on}, pages
  413--427. IEEE, 2012.

\bibitem{Meusel2014}
R.~Meusel, S.~Vigna, O.~Lehmberg, and C.~Bizer.
\newblock Graph structure in the web---revisited: A trick of the heavy tail.
\newblock In {\em WWW}, pages 427--432, 2014.

\bibitem{Meusel2015}
R.~Meusel, S.~Vigna, O.~Lehmberg, and C.~Bizer.
\newblock The graph structure in the web--analyzed on different aggregation
  levels.
\newblock {\em The Journal of Web Science}, 1(1), 2015.

\bibitem{Page1999}
L.~Page, S.~Brin, R.~Motwani, and T.~Winograd.
\newblock The pagerank citation ranking: Bringing order to the web.
\newblock Technical report, 1999.

\bibitem{Perry2013}
M.~Perry, E.~Clark, and S.~Murdoch.
\newblock The design and implementation of the tor browser [draft].
\newblock May 2015.

\bibitem{Roesner2012}
F.~Roesner, T.~Kohno, and D.~Wetherall.
\newblock Detecting and defending against third-party tracking on the web.
\newblock In {\em Proceedings of the 9th USENIX conference on Networked Systems
  Design and Implementation}, pages 12--12. USENIX Association, 2012.

\bibitem{Roosendaal2011}
A.~Roosendaal.
\newblock Facebook tracks and traces everyone: Like this!
\newblock {\em Tilburg Law School Legal Studies Research Paper Series}, (03),
  2011.

\bibitem{Schelter2016}
S.~Schelter and J.~Kunegis.
\newblock Tracking the trackers: A large-scale analysis of embedded web
  trackers.
\newblock {\em ICWSM}, 2016.

\bibitem{Spiegler2013}
S.~Spiegler.
\newblock Statistics of the common crawl corpus 2012.
\newblock Technical report, Technical report, SwiftKey, 2013.

\bibitem{EconomistDemocracy12}
{The Economist Intelligence Unit}.
\newblock Democracyindex.
\newblock 2012.

\bibitem{TheGuardian}
{The Guardian}.
\newblock Belgian court orders facebook to stop tracking non-members.
\newblock 2015.

\bibitem{TheGuardianGoogleChina}
{The Guardian}.
\newblock Google is returning to china? it never really left.
\newblock 2015.

\bibitem{TheIntercept}
{The Intercept}.
\newblock From radio to porn, british spies track web users' online identities.
\newblock 2015.

\bibitem{WorldBank15}
{The World Bank}.
\newblock Population, total.
\newblock 2015.

\bibitem{Trackography}
{Trackography}.
\newblock Meet the trackers; me and my shadow.
\newblock 2014.

\bibitem{CensusGov15}
{U.S. Census Bureau}.
\newblock U.s. trade in goods by country.
\newblock 2015.

\bibitem{RBTH2014}
P.~Vsevolod.
\newblock Russia beyond the headlines: Yandex reacts to putin comments about
  foreign influence as share price falls.
\newblock 2014.

\bibitem{Wikipedia15}
{Wikipedia}.
\newblock List of countries by english-speaking population.
\newblock 2015.

\bibitem{Yu2016}
Z.~Yu, S.~Macbeth, K.~Modi, and J.~M. Pujol.
\newblock Tracking the trackers.
\newblock In {\em Proceedings of the 25th International Conference on World
  Wide Web}, pages 121--132. International World Wide Web Conferences Steering
  Committee, 2016.

\end{thebibliography}

\end{document}